%% file: optimal_output_agreement_extensions.tex
\documentclass[a4paper, 10pt, conference]{ieeeconf}      
\IEEEoverridecommandlockouts                              
\usepackage{amsmath,amsfonts,amssymb,amscd,latexsym,enumerate,theorem,bbm,stfloats}

\usepackage{graphicx,epsfig,psfrag}
\usepackage{subfigure}
\usepackage{tikz}
\usepackage{tkz-berge}
\usepackage{float}
\usepackage{cite}
\usepackage{dsfont}
 \allowdisplaybreaks
\include{Nima-newcommands}

\usepackage[prependcaption,colorinlistoftodos]{todonotes}

{\theorembodyfont{\itshape}\newtheorem{theorem}{Theorem}}
{\theorembodyfont{\itshape}\newtheorem{proposition}[theorem]{Proposition}}
{\theorembodyfont{\itshape}\newtheorem{lemma}[theorem]{Lemma}}
{\theorembodyfont{\itshape}\newtheorem{corollary}[theorem]{Corollary}}
{\theorembodyfont{\upshape}\newtheorem{definition}[theorem]{Definition}}
{\theorembodyfont{\itshape}}
{\theorembodyfont{\upshape}\newtheorem{remark}[theorem]{Remark}}
{\theorembodyfont{\upshape}}
{\theorembodyfont{\upshape}\newtheorem{example}[theorem]{Example}}

\newcommand{\ones}{\mathds{1}}

\newcommand{\map}[2]{\ensuremath{:#1 \rightarrow #2}}
\newcommand{\part}[2]{\ensuremath{\frac{\partial #1}{\partial #2}}}

\newcommand{\ppower}[2]{\ensuremath{#1^{#2}}}
\title{\textbf{Output agreement in networks with\\
	unmatched disturbances and algebraic constraints}}

\author{Nima Monshizadeh \and Claudio De Persis
	\thanks{Nima Monshizadeh and Claudio De Persis are with the Engineering and Technology Institute, University of Groningen, The Netherlands, {\tt\small n.monshizadeh@rug.nl, c.de.persis@rug.nl}}
}

\begin{document}
\maketitle
\begin{abstract}
This paper considers a problem of output agreement in heterogeneous networks with dynamics on the nodes as well as on the edges. The control and disturbance signals entering the nodal dynamics are ``unmatched" meaning that some nodes are only subject to disturbances, and are deprived of actuating signals. To further enrich our model, we accommodate (solvable) algebraic constraints in a subset of nodal dynamics. We show that appropriate dynamic feedback controllers
achieve output agreement on a desired vector. We also investigate the case of an optimal steady-state control over the network. The proposed results are applied to a heterogeneous microgrid. 
\end{abstract}

	\section{Introduction}
	
Agreement on a certain quantity of interest plays a central role in cooperative control. The most notable instances are distributed optimization \cite{Tsitsiklis1986}, consensus \cite{OlfatiMurray1}, formation control \cite{OlfatiSaber:07}, and synchronization, see e.g.
\cite{Stan2007}, \cite{Li-consensus:99}, \cite{Trent-rob-sync:13}.

The study of output agreement/regulation problem in the presence of disturbances has been motivated by numerous applications in balancing demand and supply, power networks,  and hydraulic networks. In this framework, the demands/loads are interpreted as external disturbances affecting the network dynamics, see e.g. \cite{DePersis2013}, \cite{burger.et.al.mtns14b}, \cite{dorfler2013synchronization}.  

In this paper, we consider agents with non-identical nonlinear port-Hamiltonian dynamics; see \cite{van2014port} for more information on port-Hamiltonian systems.
The nodal dynamics is subject to constant disturbances.
In addition, we consider that a subset of nodal dynamics is governed by algebraic constraints. These constraints could be the result of mismatch in the dynamic order of the agents, or an approximation of fast subdynamics in singularly perturbed models \cite{kokotovic-singular}.
The algebraic constraints we consider here are solvable meaning that they can be expressed in terms of other state variables of the network.
However, obviously, the presence of such constraints adds to the heterogeneity of the network, and complicates the analysis.

We consider the physical coupling to be ``undamped", and given by a single integrator with a nonlinear output map.  
We first show that an equilibrium of the network, if  exists, is attractive and thus output agreement is locally achieved for the network.
Next, we include controller dynamics on the nodes to guarantee output agreement on a prescribed set point, in the presence of physical coupling and disturbance signals. Another important feature here is that we treat an {\em unmatched} control-disturbance scheme, meaning that control signals and disturbances may act on different subsets of nodes.
As a case study, we consider a heterogeneous microgrid consisting of synchronous generators,  droop-controlled inverters, and frequency dependent loads, where the goal is to guarantee a zero frequency deviation for all the nodes of the grid.

Note that the main contribution of the current manuscript is to consider simultaneously i) multivariable nonlinear nodal dynamics, ii) dynamic physical coupling, iii) algebraic constraints, and iv) unmatched disturbances in the output agreement problem. 
Our analysis is implicitly based on passivity and incremental passivity property inspired by \cite{Arcak2007}, \cite{Bai2011}, \cite{Burger2013}, \cite{burger.depersis.aut15}, \cite{schaft2013}.

The analysis of output agreement problem is carried out in Section \ref{s:analysis}, whereas the control design is treated in Section \ref{s:control}. Section \ref{s:case} is devoted to the case study of microgrids. Conclusions are provided in Section \ref{s:conc}. The formal proofs of the proposed results are collected in the appendix in Section \ref{s:app}. 

\noindent \textbf{Notation }
Apart form the standard notation, we use the following conventional notation. We use superscripts for vectors and matrices to indicate their domain of  definition. In particular, let 
$x_j$ with $j\in \calI$ be a set of vectors. Then, by \ppower{x}{i} we mean
$\ppower{x}{i}=\col(x_j)$ with $j\in \calI_{i}\subseteq \calI$. For a set of matrices, we define $\ppower{A}{i}=\bdiag(A_j)$ with $j\in\calI_i\subseteq \calI$.
We remove the superscript in case $\calI_i=\calI$.
\section{analysis}\label{s:analysis}

We define a dynamical network on a connected undirected  graph $\calG=(\calV, \calE)$. 
We partition the vertex set of $\calG$ into two distinct subsets, $\calV:=\calI= \calI_1 \cup \calI_2$.
To each vertex of $\calG$, we associate the following port-Hamiltonian types of dynamics:
\bse\label{e:node}
\begin{align}
\dot{x}_i&=(J_i-R_i)\nabla H_{n,i}(x_i)+G_i(\sigma_i +d_i)    &&i\in \calI_1\\
0&=(J_i-R_i)\nabla H_{n,i}(x_i)+G_i(\sigma_i +d_i)  &&i\in \calI_2\\
y_i&=G_i^T\nabla H_{n,i}(x_i) &&i\in \calI
\end{align}
\ese
where $x_i\in \R^n$, $J_i$ is a skew symmetric matrix, $R_i$ is a positive definite matrix, $\sigma_i\in \R^m$ amounts for the physical coupling, $d_i\in \R^m$ is a constant vector, and the Hamiltonian $H_{n,i} \map{\R^n}{R}$ is strictly convex in an open convex set $\Omega_n \subseteq \R^n$ for each $i$.


To each edge of $\calG$, we associate the following dynamics:
\bse\label{e:edge}
\begin{align}
\dot{\eta}_k&=v_k\\
\mu_k &=\nabla H_{e,k}(\eta_k)
\end{align}
\ese
where $\eta_k\in \R^m$, the Hamiltonian $H_{e,k} \map{\R^m}{\R}$ is strictly convex in an open convex set $\Omega_e \subseteq \R^m$, and $k=\seq{M}$.
The interconnection law is given by
\beq\label{e:interconnection}
v= (B^T \otimes I_m) y, \quad \sigma= -(B \otimes I_m) \mu
\eeq
where $B$ is the incidence matrix of $\calG$, $v=\col(v_k)$, $y=\col(y_i)$, and $\sigma=\col(\sigma_i)$
with $k=\seq{M}$ and $i=\seq{N}$. 

Then, the edge dynamics \eqref{e:edge}, the nodal dynamics \eqref{e:node}, and the interconnection law  \eqref{e:interconnection} can be written compactly as
\bse\label{e:system}
\begin{align}
\label{e:eta}
 \dot \eta &= (B^T \otimes I) G^T\nabla H_{n} (x)    \\
\mu &= \nabla H_{e} (\eta)  \\ 
\nonumber
\ppower{\dot x}{1}&= (\ppower{J}{1}- \ppower{R}{1}) \nabla \ppower{H_{n}}{1} (\ppower{x}{1})\\
&\qquad \quad   -\ppower{G}{1}(\ppower{B}{1} \otimes I) \nabla H_{e} (\eta) + \ppower{G}{1}\ppower{d}{1}\\
\nonumber 0&= (\ppower{J}{2}- \ppower{R}{2}) \nabla \ppower{H_{n}}{2} (\ppower{x}{2})\\ \label{e:alg} &\qquad \quad -\ppower{G}{2}(\ppower{B}{2} \otimes I) \nabla H_{e} (\eta) +  \ppower{G}{2}\ppower{d}{2}\\
y &= G^T\nabla H_{n} (x)
\end{align}
\ese
where $\ppower{B}{1}$ and $\ppower{B}{2}$ denote the submatrices obtained from $B$ by collecting the rows indexed by $\calI_1$ and $\calI_2$, respectively.  
Let $x=\col(\ppower{x}{1}, \ppower{x}{2})$ and $d=\col(\ppower{d}{1}, \ppower{d}{2})$.
Suppose that $(\barx, \bar\eta) \in (\Omega_n)^N \times (\Omega_e)^M$ is an equilibrium of system \eqref{e:system} with $\dot{\barx}=0$ and $\dot{\bar{\eta}}=0$.
Then, we have
\bigskip{}
\vspace*{-\belowdisplayshortskip}
\bse\label{e:equib}
\begin{align}
\label{e:feas-agree}  0 &= (B^T \otimes I) G^T\nabla H_{n} (\barx)    \\
\nonumber
0  &= (\ppower{J}{1}-\ppower{R}{1}) \nabla \ppower{H_{n}}{1} 
(\ppower{\barx}{1})\\
&\qquad \quad   -\ppower{G}{1}(\ppower{B}{1}\otimes I) \nabla H_{e} (\bar\eta) + \ppower{G}{1}\ppower{d}{1} \\
\nonumber
0&= (\ppower{J}{2}-\ppower{R}{2})\nabla \ppower{H_{n}}{2} (\ppower{\barx}{2})\\
&\qquad \quad  -\ppower{G}{2}(\ppower{B}{2} \otimes I) \nabla H_{e} (\bar\eta) 
+\ppower{G}{2}\ppower{d}{2}. 
\end{align}
\ese
Observe that the equation \eqref{e:feas-agree} yields an output agreement condition 
\beq\label{e:agree}
G_i^T\nabla H_{n,i} (\overline x_i) = G_j^T\nabla H_{n,j} (\overline x_j), \quad \forall i, j\in \calI.
\eeq
Hence, we obtain that $ G^T\nabla H_n(\barx)=\ones_N \otimes  y^\ast$ for some 
constant vector $y^\ast \in \R^n$.
The other two equations can be written together as
\beq\label{e:feas-0}
0=(J-R)\nabla H_n(\barx)-G (B\otimes I) \nabla H_e(\bar\eta)+Gd.
\eeq
This implies that
\beq\label{e:feas}
\ones_N \otimes y^\ast=G^T(J-R)^{-1} G((B\otimes I) \nabla H_e(\bar\eta)-d).
\eeq

In case the matrix $G$ is equal to the identity matrix, by multiplying both hand sides of \eqref{e:feas} from the left by $(\ones_N^T \otimes I_n)(J-R)$, we obtain that 
\beq\label{e:feas-I}
\sum^N_{i=1} (J_i-R_i) y^\ast=-\sum_{i=1}^N d_i.
\eeq
Hence, $y^\ast=\nabla H_{n,i}(\barx_i)$ is computed as
\beq\label{e:uniq-x}
y^\ast=-(\sum^N_{i=1} (J_i-R_i))^{-1}\sum_{i=1}^N d_i.
\eeq 
Then, noting that $\ones_N \otimes y^\ast=\nabla H_n(\barx)$, the constant vector $\bar{x}\in (\Omega_n)^N$  is unique in this case.
It is worth mentioning that in the case $n=1$, we have $J=0$, and \eqref{e:uniq-x} is simplified to
$
y^\ast=\frac {\ones^T d}{\ones^T R \ones} 
.$ 

By replacing \eqref{e:uniq-x} in \eqref{e:feas} with $G=I$, the term 
$(B\otimes I)\nabla H_e(\bar\eta)$ is explicitly computed. 
Hence, $\bar\eta \in (\Omega_e)^M$ in general is not unique.
However, in case the graph $\calG$ is acyclic, the incidence matrix $B$ has full column rank, and thus $\bar\eta$ is unique.
Note that an equilibrium $(\barx, \bar\eta)\in (\Omega_n)^N\times (\Omega_e)^M$ does not always exist, and in particular the {\em feasibility conditions} \eqref{e:agree} and \eqref{e:feas-0} must hold. The following theorem investigates stability/attractivity properties of such an equilibrium, assuming that it exists.  
\begin{theorem}\label{t:analysis}
Suppose that $(\bar{x}, \bar\eta)\in (\Omega_n)^N \times (\Omega_e)^M$ is an equilibrium of \eqref{e:system}. 
Then there exists a region of state space, which includes $(\barx, \bar\eta)$, such that any solution $(x, \eta)$ of \eqref{e:system} starting in this region asymptotically converges to an equilibrium of \eqref{e:system}, and the output agreement condition \eqref{e:agree} holds.
\end{theorem}
\textbf{Proof.}\; See Appendix. 

\medskip{}
Note that Theorem \ref{t:analysis} implies that the network \eqref{e:system} reaches an output agreement providing that there exist constant vectors $(\barx, \bar\eta)\in (\Omega_n)^N\times (\Omega_e)^M$ satisfying \eqref{e:agree}, \eqref{e:feas-0}, and thus \eqref{e:feas}. As the vector $y^\ast$ resulting from this agreement may be not the desired one, due to the dependency on the disturbance $d$, next we investigate the possibility to influence this vector by an appropriate control scheme.
\section{control}\label{s:control}
In this section, we treat certain control problems related to network dynamics \eqref{e:system}. 
 To capture the heterogeneous role of the nodes, we further partition the nodal dynamics \eqref{e:node} as
\begin{align}\label{e:node-control}
\nonumber
\dot{x}_i&=(J_i-R_i)\nabla H_{n,i}(x_i)+G_i(\sigma_i +u_i +\delta_i) &&i\in \calI_{11}\\
\nonumber
\dot{x}_i&=(J_i-R_i)\nabla H_{n,i}(x_i)+G_i(\sigma_i +\delta_i)  &&i\in \calI_{12}\\
\nonumber
0&=(J_i-R_i)\nabla H_{n,i}(x_i)+G_i(\sigma_i +u_i +\delta_i)  &&i\in \calI_{21}
\\
\nonumber
0&=(J_i-R_i)\nabla H_{n,i}(x_i)+G_i(\sigma_i +\delta_i) &&i\in \calI_{22}\\
y_i&=G_i^T\nabla H_{n,i}(x_i) \qquad  \qquad &&i\in \calI
\end{align}
where $\calI_1=\calI_{11}\cup \calI_{12}$, $\calI_2=\calI_{21}\cup \calI_{22}$, $\calI_{11}\neq \emptyset$, and $G_i$ has a full column rank for each $i$. 
Here, the $u_i$ components are treated as control signals which are applied to the nodes, and the $\delta_i$s are constant disturbance signals affecting the nodal dynamics. Note that the nodes in $\calI_{12}$ and $\calI_{22}$ are not directly controlled, and therefore our treatment here incorporates the case of an {\em unmatched} control-disturbance scheme.

Now, the overall network dynamics can be written as
\bse\label{e:system-control}
\begin{align}
\label{e:control-eta}
\dot \eta &= (B^T \otimes I) G^T\nabla H_{n} (x)    \\
\nonumber \ppower{\dot x}{11}&= (\ppower{J}{11}-\ppower{R}{11}) 
\nabla\ppower{ H_{n}}{11}(\ppower{x}{11})\\
\nonumber &\qquad\quad  -\ppower{G}{11}(\ppower{B}{11}\otimes I) \nabla H_{e} (\eta) \\
\label{e:control-11}
&\qquad\quad+ \ppower{G}{11} \ppower{u}{11}+\ppower{G}{11}\ppower{\delta}{11}\\
\nonumber
\ppower{\dot x}{12}&= (\ppower{J}{12}-\ppower{R}{12}) \nabla 
\ppower{H_{n}}{12}(\ppower{x}{12})\\ 
\label{e:control-12}
&\qquad\;-\ppower{G}{12}(\ppower{B}{12} \otimes I) \nabla H_{e} (\eta) +\ppower{G}{12}\ppower{\delta}{12}\\*
\nonumber
0&= (\ppower{J}{21}-\ppower{R}{21}) \nabla \ppower{H_{n}}{21} (\ppower{x}{21})\\  
\nonumber
&\qquad\quad-\ppower{G}{21}(\ppower{B}{21} \otimes I) \nabla H_{e} (\eta) 
\\
\label{e:alg-control1}
&\qquad\quad+ \ppower{G}{21} \ppower{u}{21}+\ppower{G}{21}\ppower{\delta}{21} \\
\nonumber
0&= (\ppower{J}{22}-\ppower{R}{22}) \nabla \ppower{H_{n}}{22} (\ppower{x}{22})\\*  
\label{e:alg-control2}
&\qquad\;-\ppower{G}{22}(\ppower{B}{22} \otimes I) \nabla H_{e} (\eta) + \ppower{G}{22}\ppower{\delta}{22} \\
y &= G^T\nabla H_{n} (x).
\end{align}
\ese

Our goal here is to design dynamic feedback controllers $\ppower{u}{11}$ and $\ppower{u}{21}$ such that output agreement \eqref{e:agree} is guaranteed for the network, for a prescribed vector $y^\ast$, in the presence of network coupling and disturbance signals. 
If such $\ppower{u}{11}$ and $\ppower{u}{21}$ exist, we say that
the output agreement problem is {\em solvable}.    
Obviously, this may not be always plausible, and by \eqref{e:system-control} 
we obtain the following feasibility condition
\bse\label{e:feas-control-0}
\begin{align}
	\ones \otimes y^\ast&= G^T\nabla H_{n} (\barx)    \\
	\nonumber
	0  &= (\ppower{J}{11}-\ppower{R}{11}) \nabla \ppower{H_{n}}{11} (\ppower{\barx}{11})\\ 
	\nonumber
	&\qquad  -\ppower{G}{11}(\ppower{B}{11}\otimes I) \nabla H_{e} (\bar\eta)\\ &\qquad+ \ppower{G}{11} \ppower{\baru}{11}+\ppower{G}{11} \ppower{\delta}{11}\\
	\nonumber
	0  &= (\ppower{J}{12}-\ppower{R}{12}) \nabla \ppower{H_{n}}{12} (\ppower{\barx}{12})\\ 
	&\quad -\ppower{G}{12}(\ppower{B}{12} \otimes I) \nabla H_{e} (\bar\eta) +\ppower{G}{12}\ppower{\delta}{12}\\
	\nonumber
	0&= (\ppower{J}{21}-\ppower{R}{21}) \nabla \ppower{H_{n}}{21} (\ppower{\barx}{21})\\ &\qquad -\ppower{G}{21}(\ppower{B}{21}\otimes I) \nabla H_{e} (\bar\eta) \\
	\nonumber
	&\qquad+\ppower{G}{21} \ppower{\baru}{21}+\ppower{G}{21} \ppower{\delta}{21} \\
	\nonumber
	0&= (\ppower{J}{22}-\ppower{R}{22}) \nabla \ppower{H_{n}}{22} (\ppower{\barx}{22})\\
	&\quad -\ppower{G}{22}(\ppower{B}{22}\otimes I) \nabla H_{e} (\bar\eta) + \ppower{G}{22}\ppower{\delta}{22} 
\end{align}
\ese
Clearly, this boils down to the following condition.

\noindent  {\textbf{Feasibility condition:}} there exist constant vectors $\barx\in (\Omega_n)^N$, $\bar{\eta}\in (\Omega_e)^M$, $\ppower{d}{11}$, $\ppower{d}{21}$ such that
\bse\label{e:feas-control}
\begin{align}
\ones \otimes y^\ast&= G^T\nabla H_{n} (\barx)    \\
\nonumber
0  &= (\ppower{J}{11}-\ppower{R}{11}) \nabla \ppower{H_{n}}{11} (\ppower{\barx}{11})\\  
&\quad\;
-\ppower{G}{11}(\ppower{B}{11} \otimes I) \nabla H_{e} (\bar\eta) + \ppower{G}{11} \ppower{d}{11}\\
\nonumber
0  &= (\ppower{J}{12}-\ppower{R}{12}) \nabla \ppower{H_{n}}{12} (\ppower{\barx}{12})\\  
&\quad\;
-\ppower{G}{12}(\ppower{B}{12} \otimes I) \nabla H_{e} (\bar\eta) +\ppower{G}{12}\ppower{\delta}{12}\\
\nonumber
0&= (\ppower{J}{21}-\ppower{R}{21}) \nabla \ppower{H_{n}}{21} (\ppower{\barx}{21})\\ 
&\quad\;
-\ppower{G}{21}(\ppower{B}{21} \otimes I) \nabla H_{e} (\bar\eta) + \ppower{G}{21} \ppower{d}{21} \\
\nonumber
0&= (\ppower{J}{22}-\ppower{R}{22}) \ppower{\nabla H_{n}}{22} (\ppower{\barx}{22}) \\ 
&\quad\; -\ppower{G}{22}(\ppower{B}{22}\otimes I) \nabla H_{e} (\bar\eta) + \ppower{G}{22}\ppower{\delta}{22} 
\end{align}
\ese
Note that we have used the fact that $\bar\eta$ is constant, and $\ppower{G}{11}$ and $\ppower{G}{22}$ are full column rank. 
%
%
%
Now, we have the following result.
\vspace*{-\belowdisplayshortskip}
\begin{theorem}\label{t:control-constant}
Consider the decentralized controller 
\bse\label{e:controller-const}
\begin{align}
\dot{\xi}_i&=y^\ast-G_i^T\nabla H_{n,i}(x_i)\\
u_i &={\xi}_i
\end{align}
\ese
with $i \in \calI_{11}\cup \calI_{21}$.
Assume that the feasibility condition \eqref{e:feas-control} holds, and let
$\xi=\col(\xi_i)$ and $\bar\xi=\col(\ppower{d}{1}, \ppower{d}{2})$.
Then, there exists a region of state space, including $(\barx, \bar\eta, \bar\xi)$, such that any solution $(x, \eta, \xi)$ of the network asymptotically converges to an equilibrium of  \eqref{e:system-control} and \eqref{e:controller-const}, in which $G_i^\ast\nabla H_{n,i}(\barx_i)=y^\ast$ for each $i\in \calV$.
\end{theorem}

\BP See Appendix.
\bre
Note that in case the controller at a node $i\in \calI_{11}$ or $i\in \calI_{21}$ does not have access to the desired output $y^\ast$, 
one can set $u_i$ to a constant, namely a nominal value, and incorporate the node $i$ in the subdynamics of \eqref{e:node-control} corresponding to the nodes indexed by $\calI_{12}$ or $\calI_{22}$, respectively. 
\ere

In Theorem \ref{t:control-constant}, the control input $u$ has been designed such that output agreement on a prescribed vector $y^\ast$ is achieved for the network. Observe that the ``steady-state" control signal $\baru=\bar\xi$ is primarily determined by the initialization of the system/controller.  Next, under the constraint of output agreement \eqref{e:agree}, we aim to minimize the following quadratic cost function
\beq\label{e:cost}
\underset{\baru}{\min} =\frac{1}{2} \sum_{i\in \calI_c} \baru_i^TQ_i\baru_i
\eeq
where $Q_i\in \R^m \times \R^m$ is a positive definite matrix for each $i$, and $\calI_c=\calI_{11}\cup \calI_{21}$. 
Note that the optimization above determines the steady-state distribution of the control effort over the agents of the network. This is particularly relevant in applications involving demand and supply balancing, including power as well as hydraulic networks; see e.g. \cite{burger.et.al.mtns14b}, \cite{DePersis2013}.

To make the analysis more concise, we restrict our attention to the case where $G_i=I$ for each $i$. Then, similar to \eqref{e:feas-I}, we obtain the following constraint
\beq\label{e:feas-optimal}
\sum^N_{i=1} (J_i-R_i) y^\ast+\sum_{i\in \calI_c} u_i+\sum_{i=1}^N \delta_i=0.
\eeq
By standard Lagrange multipliers method, the vector $\baru$ which minimizes \eqref{e:cost} subject to \eqref{e:feas-optimal} is obtained as
%
\beq\label{e:optimal}
\baru_i=Q_i^{-1} \lambda
\eeq
where $\lambda\in \R^n$ is given by 
\beq\label{e:lambda}
\lambda=-(\sum_{i\in \calI_c} Q_i^{-1}) ^{-1}(\sum^N_{i=1} (J_i-R_i) y^\ast+\sum_{i=1}^N \delta_i).
\eeq
It is easy to observe that, by \eqref{e:feas-control-0} and \eqref{e:optimal}, we obtain the following feasibility condition in this case.

\noindent \textbf{Feasibility condition with optimality:}
For a given $y^\ast\in \Omega_n$, there exists a constant vector $\bar\eta\in (\Omega_e)^M$ such that
\bse\label{e:feas-control-opt}
\begin{align}
\nonumber
0  &= (\ppower{J}{11}- \ppower{R}{11}) (\ones \otimes y^\ast)  -(\ppower{B}{11} \otimes I) \nabla H_{e} (\bar\eta)\\
&\qquad \quad + (\ppower{Q}{11})^{-1}(\ones \otimes \lambda)+\ppower{\delta}{11}\\
0  &= (\ppower{J}{12}- \ppower{R}{12})(\ones \otimes y^\ast) -(\ppower{B}{12} \otimes I) \nabla H_{e} (\bar\eta) +\ppower{\delta}{12}\\
\nonumber
0&= (\ppower{J}{21}- \ppower{R}{21}) (\ones \otimes y^\ast) -(\ppower{B}{21} \otimes I) \nabla H_{e} (\bar\eta) \\
&\qquad \quad + (\ppower{Q}{21})^{-1}(\ones \otimes \lambda)+\ppower{\delta}{21}\\
0&= (\ppower{J}{22}- \ppower{R}{22}) (\ones \otimes y^\ast)
-(\ppower{B}{22} \otimes I) \nabla H_{e} (\bar\eta) +\ppower{\delta}{22} 
\end{align}
\ese
where $\lambda$ is as in \eqref{e:lambda}.

To achieve output agreement problem with an optimal ``steady state" control input, we propose a distributed controller at the nodes. The communication among the controllers takes place over a communication  graph, say $\calG_C=(\calV_c, \calE_c)$, which is undirected and connected. 
\begin{theorem}\label{t:control-opt}
Assume that the feasibility condition \eqref{e:feas-control-opt} holds. 
Consider the distributed controller 
\bse\label{e:controller-opt}
\begin{align}
\label{e:controller-opt-xi}
\dot\xi_i&=\sum_{\{i,j\}\in \calE_c} (\xi_j-\xi_i)+Q_i^{-1}(y^\ast-\nabla H_{n,i}(x_i))\\
u_i&=Q_i^{-1}\xi_i 
\end{align}
\ese
with $i \in \calI_{11}\cup \calI_{21}$.
Let $\xi=\col(\xi_i)$, and let the constant vector $\bar\xi$ be chosen as
$\bar\xi=\ones \otimes \lambda$ where $\lambda$ is given by \eqref{e:lambda}. 
Then, there exists a region of state space, including $(\barx, \bar\eta, \bar\xi)$, such that any solution $(x, \eta, \xi)$ of the network starting in this region asymptotically converges to an equilibrium of  \eqref{e:system-control} and \eqref{e:controller-opt}, in which $\nabla H_{n,i}(\barx_i)=y^\ast$ for each $i\in \calV$. Moreover, in this region, 
$u_i$ asymptotically converges to the optimal $\baru_i$ given by \eqref{e:optimal}.
\end{theorem}
\BP See Appendix.

\section{Case study}\label{s:case}

We consider a (fairly) general heterogeneous microgrid which consists of synchronous generators, droop-controlled inverters, and frequency dependent loads. We partition the buses, i.e. the nodes of $\calG$, into three sets, namely $\calV_G$, $\calV_I$, and $\calV_{L}$, corresponding to the set of synchronous generators, inverters, and loads, respectively.

The dynamics of each synchronous generator is governed by the so-called {\em swing equation}, and is given by:
\beq\label{e:gen}
M_i \ddot{\theta_i}=-A_i\dot{\theta_i}+u_i-P_{i}+\delta_i,  \qquad i\in \calV_G,
\eeq
where 
\beq\label{e:active}
P_i=\sum_{\{i,j\}\in \calE} \Imag (Y_{ij})V_iV_j \sin(\theta_i-\theta_j)
\eeq
is the active nodal injection at node $i$. Here, $M_i>0$ is the moment of inertia, $A_i>0$ is the damping constant, $u_i$ is the local controllable power generation, and $\delta_i$ is the local load at node  $i\in V_\calG$. The value of $Y_{ij}\in \C$ is equal to the admittance of the branch $\{i,j\} \in \calE$, and $\theta_i$ is the voltage angle at node $i$. Also, $V_i$ is the voltage magnitude at node $i$, and is assumed to be constant. 

For the droop-controlled inverters, we consider the following first-order model
\beq\label{e:inverters}
A_i\dot{\theta_i}=u_i-P_i+\delta_i \;, \qquad i\in \calV_I
\eeq    
where $A_i$ is known as the droop coefficient, $u_i$ is the injection power at node (inverter) $i$, $\delta_i$ is the local load at inverter $i$, and $\dot{\theta_i}$ indicates the frequency deviation from the nominal frequency of the network, $i\in \calV_I$.
The term $P_i$ has the same expression as in \eqref{e:active}. 

As for nodal dynamics corresponding to the loads, we consider frequency dependent loads given by the first-order system 
\beq\label{e:load-frequency}
A_i\dot{\theta_i}=\delta_{i}-P_i \;, \qquad i\in \calV_{L}
\eeq
Again, here $\dot{\theta}_i$ is the frequency deviation, $A_i>0$ is the damping coefficient, 
$P_i$ is given by \eqref{e:active}, and $\delta_i$ is the constant power consumption at node $i\in \calV_{L}$.  

To write the system in a compact form, we need the following nomenclature. 
For each $i=1,2, \ldots, k$, let $\gamma_k$ be defined as $\gamma_k=(\Imag{Y_{ij}})V_iV_j$ with $\{i,j\}$ being the $k^{th}$ edge of the graph, where the edge numbers are in accordance with the incidence matrix $B$. We define the diagonal matrix $\Gamma$ as $\Gamma=\diag(\gamma_j)$, with $j=1, 2, \ldots, k$. Let the matrices $B_G$, $B_I$, and $B_L$ be obtained from $B$ by collecting the rows indexed by $\calV_G$, $V_I$, and $V_L$, respectively. 
We define the vectors and matrices $M_G$, $A_G$, $\theta_G$, and $u_G$,  as $M_G=\diag(M_i)$, $A_G=\diag(A_i)$, 
$\theta_G=\col(\theta_i)$, $u_G=\col(u_i)$, and $\delta_G=\col(\delta_i)$  where $i\in V_G$.
The vectors and matrices $A_I$, $\theta_I$, and $u_I$ are defined as 
$A_I=\diag(A_i)$, $\theta_I=\col(\theta_i)$, $u_I=\col(u_i)$, and $\delta_G=\col(\delta_i)$ with $i\in \calV_I$.
In addition, let 
$A_L=\diag(A_i)$, $\theta_{L}=\col(\theta_i)$ and $\delta_{L}=\col(\delta_{i})$ where $i \in \calV_{L}$. 
Finally, let $P=\col(P_i)$, $\theta=\col(\theta_G, \theta_I, \theta_{L})$, and $\underline{\sin}(x):=\col(\sin(x_i))$ for a given vector $x$.  
Then, it is easy to observe that the dynamics of the synchronous generators, the inverters, and the loads can be written compactly as:
\bse\label{e:compact}
\begin{align}
M_G\ddot{\theta}_G+A_G\dot{\theta}_G&=-B_G \Gamma \underline{\sin}(B^\top \theta)+u_G-\delta_G
\\
A_I\dot{\theta}_I&=-B_I \Gamma \underline{\sin}(B^\top \theta) 
+u_I-\delta_I
\\
A_L\dot{\theta}_{L}&=-B_{L} \Gamma \underline{\sin}(B^\top \theta) +\delta_L
\end{align}
\ese
Note that this is the same model as \cite{dorfler2014synchronization}, see also \cite{Zhao:15}.
By defining $\eta=B^T\theta$, $\omega_G=\dot{\theta}_G$, $\omega_I=\dot{\theta}_I$, $\omega_L=\dot{\theta}_L$, and $\dot{\theta}=\omega=\col(\omega_G, \omega_I, \omega_L)$, 
the network dynamics \eqref{e:compact}, admits the following representation
\vspace*{-\belowdisplayshortskip}
\bse\label{e:grids}
\begin{align}
\dot{\eta}&=B^T\omega\\
M_G\dot{\omega}_G+A_G\omega_G&=-B_G \Gamma \underline{\sin}(\eta)
+u_G+\delta_G
\\
A_I\omega_I&=-B_I \Gamma \underline{\sin}(\eta)+u_I+\delta_I \\
A_L\omega_{L}&=-B_{L} \Gamma \underline{\sin}(\eta)+\delta_{L}
\end{align}
\ese
Now, let $p_G=M_G\omega_G$, $H_{G}=\frac{1}{2}p_G^TM_G^{-1}p_G$,
$H_I=\frac{1}{2}\omega_I^T\omega_I$, $H_L=\frac{1}{2}w_L^Tw_L$, and $H_e=-\ones^T\Gamma \underline{\rm cos}(\eta)$.
Then,  \eqref{e:grids} can be written as
\bse\label{e:grid-H}
\begin{align}
\dot{\eta}&=B^T\nabla H_T(p)\\
\label{e:grid-H-gen}
\dot{p}_G&=-A_G\nabla H_G(p_G)-B_G \nabla H_e(\eta) 
+u_G+\delta_G
\\
0 &=-A_I\nabla H_I(\omega_I)-B_I \nabla H_e(\eta)+u_I+\delta_I \\
0&=-A_L\nabla H_L(\omega_L)-B_{L} \nabla H_e(\eta)+\delta_{L}
\end{align}
where $p=\col(p_G, \omega_I, \omega_L)$ and $H_T=H_G+H_I+H_L$. 
\ese
Note that \eqref{e:grid-H} has a similar structure/properties as \eqref{e:system-control}, with $\Omega_n=\R$, $\Omega_e=(-\frac{\pi}{2}, \frac{\pi}{2})$, and
$\calI_{12}=\emptyset$.
The primary control goal here is to achieve a zero frequency deviation for the power network. As $\nabla H_T=w$, this is in accordance with our definition of output agreement with $y^\ast=0$. Moreover, we would like to achieve an optimal steady-state distribution of the power in the sense of \eqref{e:optimal}. In this case, \eqref{e:optimal} reads as
\beq\label{e:optimal-grid}
\baru_i=q_i^{-1}\lambda
\eeq
where 
$$
\lambda=-(\sum_i q_i)^{-1} (\ones^T \delta_G+\ones^T \delta_I+\ones^T \delta_L ).
$$

Observe that the feasibility condition \eqref{e:feas-control-opt} in this case amounts for the existence of a constant vector $\bar\eta\in (-\frac{\pi}{2}, \frac{\pi}{2})^M$ such that 
\bse\label{e:grid-feas}
\begin{align}
\label{e:genrator-feas}
0&=-B_G \nabla H_e(\bar\eta)+\baru_G+\delta_G
\\
0 &=-B_I \nabla H_e(\bar\eta)+\baru_I+\delta_I \\
\label{e:feas-opt-L}
0&=-B_{L} \nabla H_e(\bar\eta)+\delta_{L}
\end{align}
\ese
where $\baru_i$ is given by \eqref{e:optimal-grid} for each $i\in \calV_G \cup \calV_I $. 
Now, assume that the feasibility condition \eqref{e:grid-feas} holds. Then, by Theorem \ref{t:control-opt},
the controller
\bse\label{e:controller-grid}
\begin{align}
\dot{\xi}_i&=\sum_{\{i,j\}\in \calE_c}(\xi_j-\xi_i)-q_i^{-1}\omega_i\\
u_i&=q_i^{-1}\xi_i, \qquad i\in \calV_G\cup \calV_I	
\end{align}
\ese
achieves zero frequency deviation , and moreover $u_i$ asymptotically converges to the optimal $\baru_i$ given by \eqref{e:optimal-grid}.

Now, consider the case where a proper subset of generators, say $\calV_{F}\subset \calV_G$, encounter some failures. In particular, assume that
$u_i$ is not appropriately actuated, and is equal to some unknown constant vector for each $i\in \calV_{F}.$ Then, for the nodes in the {\em fail mode}, subdynamics \eqref{e:grid-H-gen}  reads as
\beq\label{e:grid-fail}
\dot{p}_F=-A_F\nabla H_F(p_F)-B_F\nabla H_e(\eta)+\delta_F
\eeq
where we have used the index ``F" to distinguish the subdynamics above from the nominal subdynamics \eqref{e:grid-H-gen}.  	
Assume that there exists $\eta \in (-\frac{\pi}{2}, \frac{\pi}{2})^M$ such that \eqref{e:grid-feas} and
$$0=-B_F\nabla H_e(\bar\eta)+\delta_F$$
are satisfied. Note that \eqref{e:genrator-feas} has to be modified accordingly to exclude the faulty generators. Observe that \eqref{e:grid-fail} has the same structure as 
\eqref{e:control-12}. Then, by Theorem \ref{t:control-opt}, we conclude that the controller \eqref{e:controller-grid} achieves a zero frequency deviation, and we have optimal steady state distribution of the power, given by \eqref{e:optimal-grid}, despite the failures in the nodal dynamics $\calV_G$.

Note that, similarly, absence or failure of actuation in inverters can be incorporated in our design, as this results in a similar dynamics to that of the loads.

\section{Conclusions}\label{s:conc}

We have investigated the problem of output agreement in heterogeneous networks with port-Hamiltonian nodal dynamics, dynamic physical coupling, and algebraic constraints. We have considered the case where control and disturbance signals may act on different subset of nodes.
We have observed that the equilibrium of the network, if exists, is locally attractive, and thus output variables asymptotically converges to a same vector.  As discussed, this vector can be steered to a desired one by applying decentralized dynamic controllers at the nodes, upon the satisfaction of certain feasibility conditions imposed by the physics of the problem. We have also studied the case in which we are interested in an optimal steady-state distribution of control signals over the network.
As observed, this goal can be achieved by exploiting distributed controllers at the nodes. We have applied the proposed results on a heterogeneous microgrid. Extending the analysis to incorporate time-varying disturbances is a subject of future research.      
\section{Appendix}\label{s:app}
\noindent \textbf{Poof of Theorem \ref{t:analysis}:}
From \eqref{e:eta}, we have
\beq\label{e:eta-new}
\dot{\eta}=(\ppower{B}{1} \otimes I)^T(\ppower{G}{1})^T\nabla \ppower{H_n}{1}(\ppower{x}{1})
+(\ppower{B}{2} \otimes I)^T(\ppower{G}{2})^T\nabla \ppower{H_n}{2}(\ppower{x}{2})
\eeq
By \eqref{e:alg}, we obtain that
\begin{align*}
\dot{\eta}=&\;(\ppower{B}{1} \otimes I)^T(\ppower{G}{1})^T\nabla \ppower{H_n}{1}(\ppower{x}{1})\\
&+(\ppower{B}{2} \otimes I)^T(\ppower{G}{2})^T(\ppower{J}{2}-\ppower{R}{2})^{-1}\ppower{G}{2}\\
&\quad \cdot((\ppower{B}{2} \otimes I)\nabla {H_e}(\eta)-\ppower{d}{2})
\end{align*}
Next, we study the asymptotic behavior of the following subdynamics of \eqref{e:system}
\bse\label{e:app-subdynamics}
\begin{align}
\nonumber \dot{\eta}=&\;(\ppower{B}{1} \otimes I)^T(\ppower{G}{1})^T\nabla \ppower{H_n}{1}(\ppower{x}{1})\\
\nonumber
&+(\ppower{B}{2} \otimes I)^T(\ppower{G}{2})^T(\ppower{J}{2}-\ppower{R}{2})^{-1} \ppower{G}{2}(\ppower{B}{2} \otimes I)\nabla {H_e}(\eta)\\
&\label{e:app-subdynamics-eta} -(\ppower{B}{2} \otimes I)^T(\ppower{G}{2})^T(\ppower{J}{2}-\ppower{R}{2})^{-1}\ppower{G}{2}\ppower{d}{2})\\
\nonumber
\dot x^{(1)}=& (J^{(1)}- R^{(1)}) \nabla H_{n}^{(1)} (x^{(1)})\\  &-\ppower{G}{1}(B^{(1)} \otimes I) \nabla H_{e} (\eta) + G^{(1)}d^{(1)} 
\end{align}
\ese
Let $W_n$ and $W_e$ be defined as 
\beq\label{e:Wn}
W_n(\ppower{x}{1}, \ppower{\barx}{1})=\ppower{H_n}{1}(\ppower{x}{1})
-\ppower{H_n}{1}(\ppower{\barx}{1})-(\nabla \ppower{H_n}{1}(\ppower{\barx}{1}))^T
(\ppower{x}{1}-\ppower{\barx}{1})
\eeq
and
\beq\label{e:We}
W_e(\eta, \bar\eta)=H_e(\eta)-H_e(\bar\eta)-(\nabla H_e(\bar\eta))^T(\eta-\bar\eta)
\eeq
where $(\ppower{\barx}{1}, \bar\eta)$ is an equilibrium of \eqref{e:app-subdynamics}. Following \cite{jayawardhana-passivity}, $W_n$ identifies a positive definite map with a strict local minimum at $\ppower{x}{1}=\ppower{\barx}{1}$. Also $W_e$ defines a positive definite map with a strict local minimum at $\eta=\bar\eta$. 
Noting that $\dot{\barx}=0$, we have
\begin{align*}
\dot{W}_n=&(\nabla \ppower{H_n}{1}(\ppower{x}{1}))^T \ppower{\dot{x}}{1}
-(\nabla \ppower{H_n}{1}(\ppower{\barx}{1}))^T (\ppower{\dot{x}}{1}-\ppower{\dot{\barx}}{1})\\
=&(\nabla \ppower{H_n}{1}(\ppower{x}{1})-\nabla \ppower{H_n}{1}(\ppower{\barx}{1}))^T
(\ppower{\dot{x}}{1}-\ppower{\dot{\barx}}{1})\\
=&(\nabla \ppower{H_n}{1}(\ppower{x}{1})-\nabla \ppower{H_n}{1}(\ppower{\barx}{1}))^T\\
&\qquad \cdot (\ppower{J}{1}-\ppower{R}{1})(\nabla \ppower{H_n}{1}(\ppower{x}{1})-\nabla \ppower{H_n}{1}(\ppower{\barx}{1}))\\
&-(\nabla \ppower{H_n}{1}(\ppower{x}{1})-\nabla \ppower{H_n}{1}(\ppower{\barx}{1}))^T\\ 
&\qquad \cdot\ppower{G}{1}(\ppower{B}{1} \otimes I)(\nabla H_e(\eta)-\nabla H_e(\bar\eta))
\end{align*} 
In addition, noting that $\dot{\bar\eta}=0$ we have
\begin{align}\label{e:dWe}
\nonumber
\dot{W}_e=&(\nabla H_e(\eta))^T \dot\eta-(\nabla H_e(\bar\eta))^T (\dot\eta-\dot{\bar\eta})\\
\nonumber=&(\nabla H_e(\eta)-\nabla H_e(\bar\eta))^T (\dot\eta-\dot{\bar\eta})\\
\nonumber
=&(\nabla H_e(\eta)-\nabla H_e(\bar\eta))^T (\ppower{B}{1}\otimes I)^T \\
\nonumber 
&\qquad \cdot(\ppower{G}{1})^T
(\nabla \ppower{H_n}{1}(\ppower{x}{1})-\nabla \ppower{H_n}{1}(\ppower{\barx}{1}))\\
\nonumber
&+(\nabla H_e(\eta)-\nabla H_e(\bar\eta))^T (\ppower{B}{2}\otimes I)^T
(\ppower{G}{2})^T\\
\nonumber
&\qquad \cdot(\ppower{J}{2}-\ppower{R}{2})^{-1} \ppower{G}{2}(\ppower{B}{2} \otimes I)
(\nabla {H_e}(\eta)-\nabla {H_e}(\bar\eta))\\
\end{align}

Let $W_T:=W_n+W_e$. Then, we have
\begin{align*}
\dot{W}_T&=(\nabla \ppower{H_n}{1}(\ppower{x}{1})-\nabla \ppower{H_n}{1}(\ppower{\barx}{1}))^T
(\ppower{J}{1}-\ppower{R}{1})\\
&\nonumber \qquad \cdot(\nabla \ppower{H_n}{1}(\ppower{x}{1})-\nabla \ppower{H_n}{1}(\ppower{\barx}{1}))\\
&+(\nabla H_e(\eta)-\nabla H_e(\bar\eta))^T (\ppower{B}{2}\otimes I)^T
(\ppower{G}{2})^T
\\&\nonumber \qquad \cdot
(\ppower{J}{2}-\ppower{R}{2})^{-1}\ppower{G}{2} (\ppower{B}{2} \otimes I)
(\nabla {H_e}(\eta)-\nabla {H_e}(\bar\eta))
\end{align*} 
where we have used the fact that $\ppower{d}{1}$ and $\ppower{d}{2}$ are constant.

Now, note that for any skew-symmetric matrix $J$ and a positive definite matrix $R$, we have $-2R=(J-R)+(J-R)^T <0$, and thus $(J-R)^{-1}+(J-R)^{-T}<0$. Hence, 
$z^T (J-R)z<0$ and $z^T (J-R)^{-1}z<0$ for any nonzero vector $z$.  Therefore, we conclude that $\dot{W}_T\leq0$.

Observe that $W_T$ has a strict local minimum at 
$x=\ppower{\barx}{1}$ and $\eta=\bar\eta$, and hence one can construct a compact level set $\Omega_c \subseteq (\Omega_n)^{|\calI_1|}\times (\Omega_e)^M$ around 
$(\ppower{\barx}{1}, \bar\eta)$ which is forward invariant. This implies that on the interval of definition of a solution to system \eqref{e:system}, the variables $\ppower{x}{1}$ and $\eta$ are bounded. Hence, by \eqref{e:alg}, the variables
$\nabla \ppower{H_n}{2} (\ppower{x}{2})$ are also bounded, and a solution to \eqref{e:system} exists for all $t$.

Then by invoking LaSalle invariance principle, on the invariant set $\dot{W}_T=0$, we have 
\bse\label{e:app-analysis-converge}
\begin{align}
\label{e:x1-converge}\nabla \ppower{H_n}{1}(\ppower{x}{1})-\nabla \ppower{H_n}{1}(\ppower{\barx}{1})&=0
\\
\label{e:eta-converge} \ppower{G}{2}(\ppower{B}{2} \otimes I)(\nabla {H_e}(\eta)-\nabla {H_e}(\bar\eta))&=0.
\end{align}
\ese
Due to the strict convexity of $\ppower{H_n}{1}$, \eqref{e:x1-converge} yields 
$\ppower{x}{1}=\ppower{\barx}{1}.$ Besides, \eqref{e:app-subdynamics-eta} admits the following incremental model
\begin{align*}
\dot{\eta}=&\;(\ppower{B}{1} \otimes I)^T(\ppower{G}{1})^T
(\nabla \ppower{H_n}{1}(\ppower{x}{1})-\nabla \ppower{H_n}{1}(\ppower{\barx}{1}))\\
&+(\ppower{B}{2} \otimes I)^T(\ppower{G}{2})^T(\ppower{J}{2}-\ppower{R}{2})^{-1}\\ 
&\qquad \cdot \ppower{G}{2}(\ppower{B}{2} \otimes I)(\nabla {H_e}(\eta)-\nabla {H_e}(\bar\eta))
\end{align*}
Therefore, by \eqref{e:app-analysis-converge}, we obtain that $\eta=\tilde{\eta}$ on the invariant set for some constant vector $\tilde{\eta}$, and thus output agreement \eqref{e:agree} holds. Note that, by \eqref{e:alg}, $x_2$ asymptotically converges to a constant vector identified by
\beq\label{e:new-x2}
\nabla \ppower{H_n}{2}(\ppower{\barx}{2})=
(\ppower{J}{2}-\ppower{R}{2})^{-1}\ppower{G}{2}
((\ppower{B}{2}\otimes I)\nabla H_e(\tilde{\eta})+\ppower{d}{2})
\eeq
This completes the proof.
\EP

\noindent \textbf{Proof of Theorem \ref{t:control-constant}:}
By the algebraic equation \eqref{e:alg-control1}, the controller \eqref{e:controller-const} can be written as
\bse\label{e:controller-compact}
\begin{align}
\label{e:xi11} 
\ppower{\dot{\xi}}{11}=\;&\ones \otimes y^\ast-(\ppower{G}{11})^T\nabla \ppower{H}{11}(\ppower{x}{11})\\
\nonumber
\ppower{\dot{\xi}}{21}=\;&\ones \otimes y^\ast-(\ppower{G}{21})^T
(\ppower{J}{21}-\ppower{R}{21})^{-1}\ppower{G}{21}\\ 
&\qquad \cdot((\ppower{B}{21}\otimes I) \nabla H_e(\eta)-\ppower{\xi}{21}-\ppower{\delta}{21}) 
\\
\ppower{u}{11}=&\;\ppower{\xi}{11}\\
\ppower{u}{21}=&\;\ppower{\xi}{21}.
\end{align}
\ese
Moreover, by \eqref{e:control-eta}, \eqref{e:alg-control1}, and \eqref{e:alg-control2}, we have
\begin{align}\label{e:eta-new}
\nonumber
\dot{\eta}=&\;(\ppower{B}{11}\otimes I)^T(\ppower{G}{11})^T\nabla \ppower{H}{11}(\ppower{x}{11})\\*
\nonumber
&+(\ppower{B}{12}\otimes I)^T (\ppower{G}{12})^T\nabla \ppower{H}{12}(\ppower{x}{12})\\*
\nonumber
&+(\ppower{B}{21}\otimes I)^T (\ppower{G}{21})^T(\ppower{J}{21}-\ppower{R}{21})^{-1} \ppower{G}{21}
\\*
\nonumber
&\qquad \cdot
((\ppower{B}{21}\otimes I)\nabla H_e (\eta)-\ppower{\xi}{21}-\ppower{\delta}{21})\\*
\nonumber
&+(\ppower{B}{22}\otimes I)^T(\ppower{G}{22})^T(\ppower{J}{22}-\ppower{R}{22})^{-1}\ppower{G}{22}
\\*
&\qquad \cdot
((\ppower{B}{22}\otimes I)\nabla H_e (\eta)-\ppower{\delta}{22})
\end{align}
The equation above together with \eqref{e:control-11}, \eqref{e:control-12}, and \eqref{e:controller-compact} defines a dynamical system with ordinary differential equations, the solution of which exists and is unique. Moreover, this system admits the following incremental model.
\bse\label{e:control-increment}
\begin{align}
\nonumber
\dot{\eta}-\dot{\bar{\eta}}=&\;(\ppower{B}{11}\otimes I)^T(\ppower{G}{11})^T
(\nabla \ppower{H}{11}(\ppower{x}{11})-\nabla \ppower{H}{11}(\ppower{\barx}{11}))\\*
\nonumber &
+(\ppower{B}{12}\otimes I)^T(\ppower{G}{12})^T\nabla (\ppower{H}{12}(\ppower{x}{12})-\ppower{H}{12}(\ppower{\barx}{12}))\\*
\nonumber
&\nonumber+(\ppower{B}{21}\otimes I)^T(\ppower{G}{21})^T(\ppower{J}{21}-\ppower{R}{21})^{-1}
\ppower{G}{21}\\*
&\nonumber \qquad \cdot(\ppower{B}{21}\otimes I)(\nabla H_e (\eta)-\nabla H_e (\bar{\eta}))\\*
\nonumber
&-(\ppower{B}{21}\otimes I)^T (\ppower{G}{21})^T(\ppower{J}{21}-\ppower{R}{21})^{-1}
\ppower{G}{21}(\ppower{\xi}{21}-\ppower{\xi}{21})\\*
\nonumber &+(\ppower{B}{22}\otimes I)^T(\ppower{G}{22})^T(\ppower{J}{22}-\ppower{R}{22})^{-1}\ppower{G}{22}
\\*
\label{e:increment-eta}
&\qquad \cdot
(\ppower{B}{22}\otimes I)(\nabla H_e (\eta)-\nabla H_e(\bar{\eta}))
\end{align}
\vspace*{-\belowdisplayshortskip}
\vspace*{-\belowdisplayshortskip}
\begin{align}
\nonumber
\ppower{\dot{x}}{11}-\ppower{\dot\barx}{11}=&\;(J^{11}- R^{11})
(\nabla H_{n}^{11} (x^{11})-\nabla H_{n}^{11} (\ppower{\barx}{11})\\
\nonumber &-\ppower{G}{11}(B^{11} \otimes I) (\nabla H_{e} (\eta)-\nabla H_{e} (\bar\eta))\\ 
\label{e:x11-increment}
&+ \ppower{G}{11} (\ppower{\xi}{11}-\ppower{\bar\xi}{11})\\
\nonumber
\dot x^{12}-\dot {\barx}^{12}=&\; (J^{12}- R^{12}) (\nabla H_{n}^{12} (x^{12})
-\nabla H_{n}^{12} (\barx^{12}))\\
\label{e:x12-increment}
&-\ppower{G}{12}(B^{12} \otimes I) (\nabla H_{e} (\eta)-\nabla H_{e} (\bar\eta))\\
\label{e:xi11-increment}
\ppower{\dot{\xi}}{11}-\ppower{\dot{\bar\xi}}{11}=&\;-(\ppower{G}{11})^T
(\nabla \ppower{H}{11}(\ppower{x}{11})-\nabla \ppower{H}{11}(\ppower{\barx}{11}))\\
\nonumber
\ppower{\dot{\xi}}{21}-\ppower{\dot{\bar\xi}}{21}=&\;-(\ppower{G}{21})^T
(\ppower{J}{21}-\ppower{R}{21})^{-1}\ppower{G}{21}\\
&\nonumber \qquad \cdot (\ppower{B}{21}\otimes I) 
(\nabla H_e(\eta)-\nabla H_e(\bar\eta))\\
&+(\ppower{G}{21})^T(\ppower{J}{21}-\ppower{R}{21})^{-1}\ppower{G}{21}
(\ppower{\xi}{21}-\ppower{\bar\xi}{21}) 
\end{align}
\ese
where $\ppower{\bar{\xi}}{11}=\ppower{d}{11}-\ppower{\delta}{11}$, $\ppower{\bar{\xi}}{21}=\ppower{d}{21}-\ppower{\delta}{21}$, and constant vectors $\barx$ and $\bar{\eta}$ are such that \eqref{e:feas-control} is satisfied. Note that, due to the feasibility condition, $(\barx, \bar\eta, \baru)$ is a valid solution for \eqref{e:system-control}, where $\baru=\col(\ppower{\baru}{11}, \ppower{\baru}{21}) $, $\ppower{\baru}{11}=\ppower{\bar\xi}{11}$, and $\ppower{\baru}{21}=\ppower{\bar\xi}{21}$.

Now consider the Lyapunov function
$V=W_n+W_e+W_c$ where 
\beq\label{e:Wc}
W_c=\frac{1}{2}(\ppower{\xi}{11}-\ppower{\bar{\xi}}{11})^T(\ppower{\xi}{11}-\ppower{\bar{\xi}}{11}) 
+\frac{1}{2}(\ppower{\xi}{21}-\ppower{\bar{\xi}}{21})^T(\ppower{\xi}{21}-\ppower{\bar{\xi}}
{21})
\eeq
and $W_n$ and $W_e$ are given by \eqref{e:Wn} and \eqref{e:We}, respectively.
By the use of incremental model \eqref{e:control-increment}, we obtain 
\begin{align}\label{e:dotWn}
\nonumber
\dot{W}_n=&
(\nabla \ppower{H_n}{11}(\ppower{x}{11})-\nabla \ppower{H_n}{11}(\ppower{\barx}{11}))^T
(\ppower{\dot{x}}{11}-\ppower{\dot{\barx}}{11})\\
\nonumber
&+(\nabla \ppower{H_n}{12}(\ppower{x}{12})-\nabla \ppower{H_n}{12}(\ppower{\barx}{12}))^T
(\ppower{\dot{x}}{12}-\ppower{\dot{\barx}}{12})\\
\nonumber
=&(\nabla \ppower{H_n}{11}(\ppower{x}{11})-\nabla \ppower{H_n}{11}(\ppower{\barx}{11}))^T
(\ppower{J}{11}-\ppower{R}{11})\\
\nonumber&\qquad \cdot
(\nabla \ppower{H_n}{11}(\ppower{x}{11})-\nabla \ppower{H_n}{11}(\ppower{\barx}{11}))\\
\nonumber
&-(\nabla \ppower{H_n}{11}(\ppower{x}{11})-\nabla \ppower{H_n}{11}(\ppower{\barx}{11}))^T (\ppower{G}{11})\\
\nonumber&\qquad \cdot
(\ppower{B}{11} \otimes I)(\nabla H_e(\eta)-\nabla H_e(\bar\eta))\\
\nonumber
&+(\nabla \ppower{H_n}{11}(\ppower{x}{11})-\nabla \ppower{H_n}{11}(\ppower{\barx}{11}))^T \ppower{G}{11}(\ppower{\xi}{11}-\ppower{\bar\xi}{11})\\
\nonumber
&+(\nabla \ppower{H_n}{12}(\ppower{x}{12})-\nabla \ppower{H_n}{12}(\ppower{\barx}{12}))^T
(\ppower{J}{12}-\ppower{R}{12})\\
\nonumber&\qquad \cdot
(\nabla \ppower{H_n}{12}(\ppower{x}{12})-\nabla 
\ppower{H_n}{12}(\ppower{\barx}{12}))\\
\nonumber&-(\nabla \ppower{H_n}{12}(\ppower{x}{12})-\nabla \ppower{H_n}{12}(\ppower{\barx}{12}))^T \ppower{G}{12}\\
&\qquad \cdot
(\ppower{B}{12} \otimes I)(\nabla H_e(\eta)-\nabla H_e(\bar\eta))
\end{align} 
and
\begin{align}\label{e:dotWe}
\nonumber
\dot{W}_e=\;&(\nabla H_e(\eta)-\nabla H_e(\bar\eta))^T
(\ppower{B}{11}\otimes I)^T\\*
\nonumber&\quad \cdot
(\ppower{G}{11})^T(\nabla \ppower{H}{11}(\ppower{x}{11})-\nabla \ppower{H}{11}(\ppower{\barx}{11}))
\\*
\nonumber 
&+(\nabla H_e(\eta)-\nabla H_e(\bar\eta))^T(\ppower{B}{12}\otimes I)^T\\*
\nonumber&\quad \cdot
(\ppower{G}{12})^T(\nabla \ppower{H}{12}(\ppower{x}{12})-\ppower{H}{12}(\ppower{\barx}{12}))\\*
\nonumber
&+(\nabla H_e(\eta)-\nabla H_e(\bar\eta))^T(\ppower{B}{21}\otimes I)^T(\ppower{G}{21})^T\\*
\nonumber&\quad \cdot
 (\ppower{J}{21}-\ppower{R}{21})^{-1}\ppower{G}{21}
(\ppower{B}{21}\otimes I)(\nabla H_e (\eta)-\nabla H_e (\bar{\eta}))\\
\nonumber
&-(\nabla H_e(\eta)-\nabla H_e(\bar\eta))^T(\ppower{B}{21}\otimes I)^T(\ppower{G}{21})^T\\*
\nonumber&\quad \cdot
(\ppower{J}{21}-\ppower{R}{21})^{-1}
\ppower{G}{21}(\ppower{\xi}{21}-\ppower{\bar\xi}{21})\\*
\nonumber&+(\nabla H_e(\eta)-\nabla H_e(\bar\eta))^T(\ppower{B}{22}\otimes I)^T(\ppower{G}{22})^T\\*
&\quad \cdot
(\ppower{J}{22}-\ppower{R}{22})^{-1}\ppower{G}{22}
(\ppower{B}{22}\otimes I)(\nabla H_e (\eta)-\nabla H_e(\bar{\eta}))
\end{align}
%
Then, $\dot{W}_c$ is computed as
\begin{align*}
\dot{W}_c=&-(\ppower{\xi}{11}-\ppower{\bar\xi}{11})^T
(\ppower{G}{11})^T
(\nabla \ppower{H}{11}(\ppower{x}{11})-\nabla \ppower{H}{11}(\ppower{\barx}{11}))\\
& -(\ppower{\xi}{21}-\ppower{\bar\xi}{21})^T(\ppower{G}{21})^T
(\ppower{J}{21}-\ppower{R}{21})^{-1}\ppower{G}{21}\\
\nonumber&\quad \cdot
(\ppower{B}{21}\otimes I) 
(\nabla H_e(\eta)-\nabla H_e(\bar\eta))\\
& +(\ppower{\xi}{21}-\ppower{\bar\xi}{21})^T(\ppower{G}{21})^T(\ppower{J}{21}-\ppower{R}{21})^{-1}\ppower{G}{21}
(\ppower{\xi}{21}-\ppower{\bar\xi}{21}) 
\end{align*}
Hence, we have
\begin{align*}
\dot{V}=&\;\dot{W}_n+\dot{W}_e+\dot{W}_c\\
=&(\nabla \ppower{H_n}{11}(\ppower{x}{11})-\nabla \ppower{H_n}{11}(\ppower{\barx}{11}))^T
(\ppower{J}{11}-\ppower{R}{11})\\
\nonumber&\quad \cdot
(\nabla \ppower{H_n}{11}(\ppower{x}{11})-\nabla \ppower{H_n}{11}(\ppower{\barx}{11}))\\
&+(\nabla \ppower{H_n}{12}(\ppower{x}{12})-\nabla \ppower{H_n}{12}(\ppower{\barx}{12}))^T
(\ppower{J}{12}-\ppower{R}{12})\\
\nonumber&\quad \cdot
(\nabla \ppower{H_n}{12}(\ppower{x}{12})-\nabla 
\ppower{H_n}{12}(\ppower{\barx}{12}))\\
&+(\nabla H_e(\eta)-\nabla H_e(\bar\eta))^T(\ppower{B}{22}\otimes I)^T
(\ppower{G}{22})^T\\
\nonumber&\quad \cdot
(\ppower{J}{22}-\ppower{R}{22})^{-1}\ppower{G}{22}
(\ppower{B}{22}\otimes I)(\nabla H_e (\eta)-\nabla H_e(\bar{\eta}))\\
&+z^T(\ppower{J}{21}-\ppower{R}{21})^{-1}z
\end{align*}
where
$$
z=\ppower{G}{21}(\ppower{B}{21}\otimes I)(\nabla H_e (\eta)-\nabla H_e (\bar{\eta}))
-\ppower{G}{21}(\ppower{\xi}{21}-\ppower{\bar\xi}{21})
$$
Consequently, $\dot{V}\leq0$. 
Note that $$(\ppower{x}{11}, \ppower{x}{12}, \eta, \ppower{\xi}{11}, \ppower{\xi}{21})=
(\ppower{\barx}{11}, \ppower{\barx}{12}, \bar\eta, \ppower{\bar\xi}{11}, \ppower{\bar\xi}{21})$$ is a strict local minimum of $V$, and thus one can construct a compact level set 
around this point
which is forward invariant. This implies that on the interval of definition of a solution to system \eqref{e:system}, the variables 
$\ppower{x}{11}$, $\ppower{x}{12}$, $\eta$, $\ppower{\xi}{11}$, and $\ppower{\xi}{21}$ are bounded. Therefore, by \eqref{e:alg-control1} and \eqref{e:alg-control2}, the variables $\nabla \ppower{H_n}{21}(\ppower{x}{21})$, $\nabla \ppower{H_n}{22}(\ppower{x}{21})$ are also bounded, and a solution to \eqref{e:system-control} exists for all $t$.
Now by invoking LaSalle invariance principle, one the invariant set $\dot{V}=0$, we have 
\begin{align*}
\nabla \ppower{H}{11}(\ppower{x}{11})&=\nabla \ppower{H}{11}(\ppower{\barx}{11}),
\nabla \ppower{H}{12}(\ppower{x}{12})=\nabla \ppower{H}{12}(\ppower{\barx}{12}),\\
\ppower{G}{22}&(\ppower{B}{22}\otimes I)(\nabla H_e (\eta)-\nabla H_e (\bar{\eta}))=0
\end{align*}
 and $$\ppower{G}{21}(\ppower{B}{21}\otimes I)(\nabla H_e (\eta)-\nabla H_e (\bar{\eta}))
-\ppower{G}{21}(\ppower{\xi}{21}-\ppower{\bar\xi}{21})=0$$ Hence, by \eqref{e:increment-eta}, we obtain that $\dot{\eta}=\dot{\bar{\eta}}=0$. In addition, by \eqref{e:xi11-increment}, we have $\dot{\xi}=\dot{\bar{\xi}}=0$. Consequently, by \eqref{e:xi11}, we obtain that $(\ppower{G}{11})^T\nabla  \ppower{H}{11}(\ppower{x}{11})=
(\ppower{G}{11})^T\nabla  \ppower{H}{11}(\ppower{\barx}{11})=\ones \otimes y^\ast$ on the invariant set. This together with $\dot{\eta}=0$ implies that $y_i=y^\ast$ for each $i\in \calV$. Again note that, by \eqref{e:alg-control1}-\eqref{e:alg-control2}, $\ppower{x}{21}$ and $\ppower{x}{22}$ asymptotically 
converge to constant vectors $\ppower{\barx}{21}$ and $\ppower{\barx}{22}$ with similar expressions as in \eqref{e:new-x2}, where the superscripts are modified accordingly. This completes the proof.
\EP

\noindent \textbf{Proof of Theorem \ref{t:control-opt}:}
The controller \eqref{e:controller-opt} can be written in compact as
\bse
\begin{align}
\nonumber
\bbm
\ppower{\dot{\xi}}{11}\\
\ppower{\dot{\xi}}{21}
\ebm=&
-(L_c \otimes I) 
\bbm
\ppower{\xi}{11}\\
\ppower{\xi}{21}
\ebm\\
\label{e:xi11-opt} 
&\qquad+Q^{-1}
\bbm
\ones \otimes y^\ast-\nabla \ppower{H}{11}(\ppower{x}{11})\\
\ones \otimes y^\ast-\nabla \ppower{H}{21}(\ppower{x}{21})
\ebm\\
\bbm
\ppower{u}{11}\\
\ppower{u}{21}
\ebm
=&Q^{-1}\bbm
\ppower{\xi}{11}\\
\ppower{\xi}{21}
\ebm
\end{align}
\ese 
where $L_c$ denotes the Laplacian matrix of $\calG_c$, $Q=\bdiag(Q_i)$ with
$i\in \calI_c=\calI_{11}\cup\calI_{21}$, 
and in this case
\begin{align*}
\nabla \ppower{H}{21}(\ppower{x}{21})=&(\ppower{J}{21}-\ppower{R}{21})^{-1}
\\
&\quad \cdot ((\ppower{B}{21}\otimes I) \nabla H_e(\eta)-
(\ppower{Q}{21})^{-1}
\ppower{\xi}{21}
-\ppower{\delta}{21})
\end{align*}
The controller above admits the following incremental model
\bse
\begin{align}
\nonumber
\bbm
\ppower{\dot{\xi}}{11}-\ppower{\dot{\bar\xi}}{11}\\
\ppower{\dot{\xi}}{21}-\ppower{\dot{\bar\xi}}{21}
\ebm=&
-(L_c \otimes I) 
\bbm
\ppower{\xi}{11}-\ppower{\bar\xi}{11}\\
\ppower{\xi}{21}-\ppower{\bar\xi}{21}
\ebm\\
\label{e:xi11-increment-opt} 
&\quad
-Q^{-1}
\bbm
\nabla \ppower{H}{11}(\ppower{x}{11})-\nabla \ppower{H}{11}(\ppower{\barx}{11})\\
\nabla \ppower{H}{21}(\ppower{x}{21})-\nabla \ppower{H}{11}(\ppower{\barx}{21})
\ebm
\\
\bbm
\ppower{u}{11}-\ppower{\baru}{11}\\
\ppower{u}{21}-\ppower{\baru}{21}
\ebm=&\; Q^{-1}
\bbm
\ppower{\xi}{11}-\ppower{\bar\xi}{11}\\
\ppower{\xi}{21}-\ppower{\bar\xi}{21}
\ebm
\end{align}
\ese 
where
\begin{align*}
\nonumber
\nabla\ppower{H}{21}&(\ppower{x}{21})-\nabla\ppower{H}{21}(\ppower{\barx}{21})
\\ 
&=(\ppower{J}{21}-\ppower{R}{21})^{-1}
(\ppower{B}{21}\otimes I)
 (\nabla H_e(\eta)-\nabla H_e(\bar{\eta}))\\
&\qquad -(\ppower{J}{21}-\ppower{R}{21})^{-1}
(\ppower{Q}{21})^{-1}
(\ppower{\xi}{21}-
\ppower{\bar\xi}{21})
\end{align*}
The incremental system dynamics in this case is given by
\bse\label{e:control-increment-opt}
\begin{align}
\nonumber
\dot{\eta}-\dot{\bar{\eta}}=&\;(\ppower{B}{11}\otimes I)^T
(\nabla \ppower{H}{11}(\ppower{x}{11})-\nabla \ppower{H}{11}(\ppower{\barx}{11}))\\
\nonumber
&
+(\ppower{B}{12}\otimes I)^T(\nabla \ppower{H}{12}(\ppower{x}{12})-\nabla\ppower{H}{12}(\ppower{\barx}{12}))\\
\nonumber
&\nonumber+(\ppower{B}{21}\otimes I)^T(\ppower{J}{21}-\ppower{R}{21})^{-1}
(\ppower{B}{21}\otimes I)\\
&\nonumber \qquad \quad \cdot(\nabla H_e (\eta)-\nabla H_e (\bar{\eta}))\\
\nonumber
&-(\ppower{B}{21}\otimes I)^T (\ppower{J}{21}-\ppower{R}{21})^{-1}
(\ppower{Q}{21})^{-1}(\ppower{\xi}{21}-\ppower{\xi}{21} )\\
\nonumber
&+(\ppower{B}{22}\otimes I)^T(\ppower{J}{22}-\ppower{R}{22})^{-1}
(\ppower{B}{22}\otimes I)\\
\label{e:increment-eta-opt}
&\qquad \quad \cdot
(\nabla H_e (\eta)-\nabla H_e(\bar{\eta}))
\end{align}
\begin{align}
\nonumber
\ppower{\dot{x}}{11}-\dot {\barx}^{11}=&\; (J^{11}- R^{11})
(\nabla H_{n}^{11} (x^{11})-\nabla H_{n}^{11} (\barx^{11}))\\
\nonumber
& -(B^{11} \otimes I) (\nabla H_{e} (\eta)-\nabla H_{e} (\bar\eta)) 
\\&
\label{e:x11-increment}
+(\ppower{Q}{11})^{-1}(\ppower{\xi}{11}-\ppower{\bar\xi}{11})\\
\nonumber
\dot x^{12}-\dot {\barx}^{12}  =&\;(J^{12}- R^{12}) (\nabla H_{n}^{12} (x^{12})
-\nabla H_{n}^{12} (\barx^{12}))\\
\label{e:x12-increment}
&-(B^{12} \otimes I) (\nabla H_{e} (\eta)-\nabla H_{e} (\bar\eta))
\end{align}
\ese
Now, let 
\begin{align}\label{e:app-xibar}
\ppower{\bar\xi}{11}&=\ones \otimes \lambda,  
\qquad \quad
\ppower{\bar\xi}{21}=\ones \otimes \lambda
\end{align}
where $\lambda$ is given by \eqref{e:lambda}.
By \eqref{e:app-xibar}, we have $\baru_i=Q_i^{-1}\lambda$
which coincides with $\baru_i$ given by \eqref{e:optimal}.
Hence, by \eqref{e:feas-control-opt}, it is easy to observe that $(\barx, \bar\eta, \bar{\xi})$ defines a valid solution to \eqref{e:system-control}.
%

Now consider again the Lyapunov function
$
V=W_n+W_e+W_c
$
where $W_n$, $W_e$, and $W_c$ are given by \eqref{e:Wn}, \eqref{e:We}, and \eqref{e:Wc}, receptively. Then it is  straightforward to investigate that 
\begin{align*}
\dot{V}=&\;(\nabla \ppower{H_n}{11}(\ppower{x}{11})-\nabla \ppower{H_n}{11}(\ppower{\barx}{11}))^T
(\ppower{J}{11}-\ppower{R}{11})\\
&\qquad \cdot
(\nabla \ppower{H_n}{11}(\ppower{x}{11})-\nabla \ppower{H_n}{11}(\ppower{\barx}{11}))\\
&+(\nabla \ppower{H_n}{12}(\ppower{x}{12})-\nabla \ppower{H_n}{12}(\ppower{\barx}{12}))^T
(\ppower{J}{12}-\ppower{R}{12})\\
&\qquad \cdot
(\nabla \ppower{H_n}{12}(\ppower{x}{12})-\nabla 
\ppower{H_n}{12}(\ppower{\barx}{12}))\\
&+(\nabla H_e(\eta)-\nabla H_e(\bar\eta))^T(\ppower{B}{22}\otimes I)^T(\ppower{J}{22}-\ppower{R}{22})^{-1}\\
&\qquad \cdot
(\ppower{B}{22}\otimes I)(\nabla H_e (\eta)-\nabla H_e(\bar{\eta}))
\\
&-\tilde{\xi}^T (L_c\otimes I) \tilde{\xi}+z^T(\ppower{J}{21}-\ppower{R}{21})^{-1}z
\end{align*}
where
$
\tilde{\xi}=\col(\ppower{\xi}{11}-\ppower{\bar\xi}{11},    
 \ppower{\xi}{21}-\ppower{\bar\xi}{21})$ and
\begin{align*}
z=(\ppower{B}{21}\otimes I)(\nabla H_e (\eta)-\nabla H_e (\bar{\eta}))
-(\ppower{Q}{21})^{-1}(\ppower{\xi}{21}-\ppower{\bar\xi}{21})
\end{align*}
Hence, we obtain that $\dot{V}\leq 0$. 
Note that boundedness, existence, and uniqueness of solution is guaranteed in the same vein as in the proof of Theorem \ref{t:control-constant}. 
Now by constructing a forward invariant compact level set around
$(\ppower{\barx}{11}, \ppower{\barx}{12}, \bar\eta, \ppower{\bar\xi}{11}, \ppower{\bar\xi}{21})$, and 
invoking LaSalle invariance principle, on the invariant set we have
\bse\label{e:opt-inv}
\begin{align}
\label{e:app-x11-inv}\nabla \ppower{H}{11}(\ppower{x}{11})-\nabla \ppower{H}{11}(\ppower{\barx}{11})&=0\\
\nabla \ppower{H}{12}(\ppower{x}{12})-\nabla \ppower{H}{12}(\ppower{\barx}{12})&=0\\
(\ppower{B}{22}\otimes I)(\nabla H_e (\eta)-\nabla H_e (\bar{\eta}))&=0\\
\label{e:opt-inv-xi}
(L_c\otimes I)\tilde{\xi}&=0\\
(\ppower{B}{21}\otimes I)(\nabla H_e (\eta)-\nabla H_e (\bar{\eta}))&
-(\ppower{Q}{21})^{-1}(\ppower{\xi}{21}-\ppower{\bar\xi}{21})=0
\end{align}
\ese
Therefore, by \eqref{e:increment-eta-opt}, we have $\dot{\eta}=\dot{\bar{\eta}}=0$. 
Moreover, by \eqref{e:xi11-increment-opt}, \eqref{e:app-x11-inv}, and \eqref{e:opt-inv-xi},
we obtain that $\ppower{\dot{\xi}}{11}=\ppower{\dot{\bar\xi}}{11}=0$ on the invariant set. In addition, \eqref{e:opt-inv-xi} implies that
$\tilde{\xi}=\ones \otimes \alpha$ for some vector $\alpha$. 
Replacing this into \eqref{e:xi11-opt} yields    
$\nabla  \ppower{H}{11}(\ppower{x}{11})=\nabla \ppower{H}{11}(\ppower{\barx}{11})=\ones \otimes y^\ast.$ This together with $\dot{\eta}=0$ results in $y_i=y^\ast$ for each $i\in \calV$.
Note that on the invariant set, $\ppower{\xi}{11}=\ppower{\bar\xi}{11}+\ones \otimes \alpha
=\ones \otimes (\alpha+\lambda)$. Similarly, we have
$\ppower{\xi}{21}=\ones \otimes (\alpha+\lambda)$, and, hence the system dynamics takes the form
\bse
\nonumber
\begin{align*}
0  &= (J^{11}- R^{11}) (\ones \otimes y^\ast)  -(B^{11} \otimes I) \nabla H_{e} (\tilde\eta) 
\\
&\qquad \quad +(\ppower{Q}{11})^{-1}\ones \otimes
(\lambda+ \alpha)+\ppower{\delta}{11}\\
0  &= (J^{12}- R^{12})(\ones \otimes y^\ast) -(B^{12} \otimes I) \nabla H_{e} (\tilde\eta) +\ppower{\delta}{12}\\
0&= (J^{21}- R^{21}) (\ones \otimes y^\ast) -(B^{21} \otimes I) \nabla H_{e} (\tilde\eta)\\
&\qquad \quad+(\ppower{Q}{21})^{-1}\ones \otimes (\lambda+ \alpha)+\ppower{\delta}{21}\\
0&= (J^{22}- R^{22}) (\ones \otimes y^\ast)
-(B^{22} \otimes I) \nabla H_{e} (\tilde\eta) +\ppower{\delta}{22} 
\end{align*}
\ese
where $\tilde{\eta}$ is a constant vector. 
By multiplying each of the equalities above from the left by $\ones^T \otimes I$ and taking the sum over all the resulting equalities, we conclude that  
$$
\lambda+\alpha=-(\sum_{i\in \calI_c} Q_i^{-1}) ^{-1}(\sum^N_{i=1} (J_i-R_i) y^\ast+\sum_{i=1}^N \delta_i)
$$
By comparing the equality above to \eqref{e:lambda}, we obtain that 
$\alpha=0$. Consequently, on the invariant set $u_i$ is equal to the optimal $\baru_i$ given by \eqref{e:optimal}. 
\EP
\bibliographystyle{plain}
\bibliography{ref}
\end{document}

%% file: Nima-newcommands.tex
\DeclareMathOperator{\col}{col}
\DeclareMathOperator{\diag}{diag}
\DeclareMathOperator{\bdiag}{blockdiag}

\DeclareMathOperator{\Imag}{Im}

\let\leq\leqslant

\let\emptyset\varnothing

\newcommand{\calE}{\ensuremath{\mathcal{E}}}

\newcommand{\calG}{\ensuremath{\mathcal{G}}}

\newcommand{\calI}{\ensuremath{\mathcal{I}}}

\newcommand{\calV}{\ensuremath{\mathcal{V}}}

\newcommand{\baru}{\ensuremath{\bar{u}}}

\newcommand{\barx}{\ensuremath{\bar{x}}}

\newcommand{\bmat}{\begin{matrix}}
\newcommand{\emat}{\end{matrix}}
\newcommand{\bbm}{\begin{bmatrix}}
\newcommand{\ebm}{\end{bmatrix}}
\newcommand{\bpm}{\begin{pmatrix}}
\newcommand{\epm}{\end{pmatrix}}
\newcommand{\bse}{\begin{subequations}}
\newcommand{\ese}{\end{subequations}}
\newcommand{\beq}{\begin{equation}}
\newcommand{\eeq}{\end{equation}}
\newcommand{\ben}{\begin{enumerate}}
\newcommand{\een}{\end{enumerate}}
\newcommand{\beni}{\renewcommand{\labelenumi}{\roman{enumi}.}
\renewcommand{\theenumi}{\roman{enumi}}\begin{enumerate}}
\newcommand{\eeni}{\end{enumerate}\renewcommand{\labelenumi}{\arabic{enumi}.}
\renewcommand{\theenumi}{\arabic{enumi}}}
\newcommand{\bena}{\renewcommand{\labelenumi}{\alpha{enumi}.}
\renewcommand{\theenumi}{\alpha{enumi}}\begin{enumerate}}
\newcommand{\eena}{\end{enumerate}\renewcommand{\labelenumi}{\arabic{enumi}.}
\renewcommand{\theenumi}{\arabic{enumi}}}
\newcommand{\bit}{\begin{itemize}}
\newcommand{\eit}{\end{itemize}}
\newcommand{\bthe}{\begin{theorem}}
\newcommand{\ethe}{\end{theorem}}
\newcommand{\blem}{\begin{lemma}}
\newcommand{\elem}{\end{lemma}}
\newcommand{\bprop}{\begin{proposition}}
\newcommand{\eprop}{\end{proposition}}
\newcommand{\bex}{\begin{example}}
\newcommand{\eex}{\end{example}}
\newcommand{\bas}{\begin{assumption}}
\newcommand{\eas}{\end{assumption}}
\newcommand{\bre}{\begin{remark}}
\newcommand{\ere}{\end{remark}}
\newcommand{\bcor}{\begin{corollary}}
\newcommand{\ecor}{\end{corollary}}
\newcommand{\bdfn}{\begin{definition}}
\newcommand{\edfn}{\end{definition}}

\newcommand{\seq}[1]{\ensuremath{1,2,\ldots,#1}}

\newcommand{\R}{\ensuremath{\mathbb R}}
\newcommand{\C}{\ensuremath{\mathbb C}}

\newcommand{\BP}{\noindent{\bf Proof. }}
\newcommand{\EP}{\hspace*{\fill} $\blacksquare$\bigskip\noindent}

%% file: optimal_output_agreement_extensions.bbl
\begin{thebibliography}{10}

\bibitem{Arcak2007}
M.~Arcak.
\newblock Passivity as a design tool for group coordination.
\newblock {\em IEEE Transactions on Automatic Control}, 52(8):1380--1390, 2007.

\bibitem{Bai2011}
H.~Bai, M.~Arcak, and J.~Wen.
\newblock {\em Cooperative control design: {A} systematic, passivity--based
  approach}.
\newblock Springer, New York, NY, 2011.

\bibitem{burger.depersis.aut15}
M.~B{\"u}rger and C.~{De Persis}.
\newblock Dynamic coupling design for nonlinear output agreement and
  time-varying flow control.
\newblock {\em Automatica}, 51:210--222, 2015.

\bibitem{burger.et.al.mtns14b}
M.~B{\"u}rger, C.~{De Persis}, and S.~Trip.
\newblock An internal model approach to (optimal) frequency regulation in power
  grids.
\newblock In {\em Proc. of the 21th International Symposium on Mathematical
  Theory of Networks and Systems (MTNS)}, pages 577--583, Groningen, the
  Netherlands, 2014.

\bibitem{Burger2013}
M.~B{\"u}rger, D.~Zelazo, and F.~Allg{\"o}wer.
\newblock Duality and network theory in passivity-based cooperative control.
\newblock {\em Automatica}, 50(8):2051--2061, 2014.

\bibitem{DePersis2013}
C.~{De~Persis}.
\newblock Balancing time-varying demand-supply in distribution networks: an
  internal model approach.
\newblock In {\em Proc. of the 12th European Control Conference (ECC)}, pages
  748 -- 753, Zurich, Switzerland, 2013.

\bibitem{dorfler2013synchronization}
F.~D{\"o}rfler, M.~Chertkov, and F.~Bullo.
\newblock Synchronization in complex oscillator networks and smart grids.
\newblock {\em Proceedings of the National Academy of Sciences},
  110(6):2005--2010, 2013.

\bibitem{dorfler2014synchronization}
Florian D{\"o}rfler and Francesco Bullo.
\newblock Synchronization in complex networks of phase oscillators: A survey.
\newblock {\em Automatica}, 50(6):1539--1564, 2014.

\bibitem{jayawardhana-passivity}
B.~Jayawardhana, R.~Ortega, E.~Garc{\'\i}a-Canseco, and F.~Castanos.
\newblock Passivity of nonlinear incremental systems: Application to pi
  stabilization of nonlinear rlc circuits.
\newblock {\em Systems \& control letters}, 56(9):618--622, 2007.

\bibitem{kokotovic-singular}
P.V. Kokotovic, R.E. O'malley, and P.~Sannuti.
\newblock Singular perturbations and order reduction in control theory? an
  overview.
\newblock {\em Automatica}, 12(2):123--132, 1976.

\bibitem{Li-consensus:99}
Z.-K. Li, Z.-S. Duan, G.-R. Chen, and L.~Huang.
\newblock {Consensus of multiagent systems and synchronization of complex
  networks: a unified viewpoint}.
\newblock {\em {IEEE Transactions on Circuits and Systems}},
  {57}(1):{213--224}, {2010}.

\bibitem{OlfatiSaber:07}
R.~Olfati-Saber, J.~A. Fax, and R.M. Murray.
\newblock {Consensus and cooperation in networked multi-agent systems}.
\newblock In {\em {Proceedings of the IEEE}}, volume~{95}, pages {215--233},
  {2007}.

\bibitem{OlfatiMurray1}
R.~Olfati-Saber and R.M. Murray.
\newblock Consensus protocols for networks of dynamical systems.
\newblock In {\em Proceedings of American Control Conference}, volume~2, pages
  951--956, 2003.

\bibitem{Stan2007}
G.-B. Stan and R.~Sepulchre.
\newblock Analysis of interconnected oscillators by dissipativity theory.
\newblock {\em IEEE Transactions on Automatic Control}, 52(2):256 -- 270, 2007.

\bibitem{Trent-rob-sync:13}
H.~L. Trentelman, K.~Takaba, and N.~Monshizadeh.
\newblock Robust synchronization of uncertain linear multi-agent systems.
\newblock {\em IEEE Transactions on Automatic Control}, 58(6):1511--1523, 2013.

\bibitem{Tsitsiklis1986}
J.~Tsitsiklis, D.~Bertsekas, and M.~Athans.
\newblock Distributed asynchronous deterministic and stochastic gradient
  optimization algorithms.
\newblock {\em IEEE Transactions on Automatic Control}, 31(9):803--812, 1986.

\bibitem{schaft2013}
A.~van~der Schaft and B.~Maschke.
\newblock Port-{H}amiltonian systems on graphs.
\newblock {\em SIAM Journal on Control and Optimization}, 51(2):906--937, 2013.

\bibitem{van2014port}
Arjan van~der Schaft and Dimitri Jeltsema.
\newblock {\em Port-Hamiltonian Systems Theory: An Introductory Overview}.
\newblock Now Publishers Incorporated, 2014.

\bibitem{Zhao:15}
C.~Zhao, E.~Mallada, and F.~D{\"o}rfler.
\newblock Distributed frequency control for stability and economic dispatch in
  power networks.
\newblock In {\em Proceedings of American Control Conference}, 2015.
\newblock Submitted.

\end{thebibliography}
